# Simultaneous suppression of scattering and aberration for ultra-high resolution imaging deep within scattering media

Sungsam Kang, Seungwon Jeong, Pilsung Kang, Taeseok D. Yang, Joonmo Ahn, Kyungdeok Song, and Wonshik Choi[*]

*Center for Molecular Spectroscopy and Dynamics, Institute for Basic Science, Seoul 02841, Korea*
*Department of Physics, Korea University, Seoul 02855, Korea*

Abstract: Thick biological tissues give rise to not only the scattering of incoming light waves, but also aberrations of the remaining unscattered waves. Due to the inability of existing optical imaging methodologies to overcome both of these problems simultaneously, imaging depth at the sub-micron spatial resolution has remained extremely shallow. Here we present an experimental approach for identifying and eliminating aberrations even in the presence of strong multiple light scattering. For time-gated complex-field maps of reflected waves taken over various illumination channels, we identify two sets of aberration correction maps, one for the illumination path and one for the reflection path, that can preferentially accumulate the unscattered signal waves over the multiple-scattered waves. By performing closed-loop optimization for forward and phase-conjugation processes, we demonstrated a spatial resolution of 600 nm up to the unprecedented imaging depth of 7 scattering mean free paths.

Reaching the ultimate diffraction-limit spatial resolution, which is approximately half the wavelength of the light source, has been a challenging task with imaging targets embedded deep within scattering media, such as biological tissues. Multiple scattering events attenuate light waves that preserve the original incidence momenta and generate multiply scattered waves, which act as strong background noise. As target depth is increased, these combined effects lead to the exponential decrease of the signal to noise ratio (SNR)[1]. Sub-micron scales of important biological reactions occurring inside living tissues have been out of reach as a consequence, and optical microscopy was unable to effectively support the investigation of the early stages of disease progression and the study of nervous systems.

When considering a target spatial resolution close to the ultimate diffraction limit, the attenuation of SNR by multiple light scattering is not the only problem. In fact, the so-called specimen-induced aberration is an equally important issue to address[2]. The signal waves that preserve original incidence momenta are not only attenuated in their intensity by the multiple light scattering, but their phases are also retarded due to the heterogeneity of the medium. These phase retardations of the signal waves vary depending on the propagation angle, and the retardation takes place for both the incident and returning paths[3]. These angle-dependent phase retardations cause the distortion of the reconstructed object image, and make them the main source of specimen-induced aberration. They also hinder the proper accumulation of signal waves in the image reconstruction stage and cause a further reduction in SNR in addition to that caused by multiple light scattering. For example, the specimen-induced aberrations of typical biological tissues with thicknesses of a few scattering mean free paths (MFPs) can attenuate the single scattering intensity of object images by hundreds of times. This detrimental aberration effect is much more pronounced for high-resolution imaging, as waves propagating at large incidence angles retaining high-spatial frequency information tend to pass through effectively longer paths and are thus more likely to experience large phase retardations. The real challenge of these aberrations when imaging targets in scattering media is that they are extremely difficult to identify in the presence of strong multiple light scattering.

In the past numerous attempts have tried to deal with either scattering or aberration individually, but few studies have been able to resolve both problems simultaneously. The methods for dealing with scattering include the use of temporal and/or confocal gating for the selective collection of single-scattered waves[4-9]. But the existence of specimen-induced aberrations easily undermines these gating operations. Using eigenchannels to better accumulate signal waves has been attempted, but does not guarantee aberration compensation[10,11]. In our previous work, we also reported a method termed collective accumulation of single scattering (CASS) microscopy[12], which combined both time-gated detection and spatial input-output correlation. CASS microscopy was used to preferentially accumulate single-scattered waves, which are the waves scattered only once by the target object, but not at all by the medium. This has resulted in a dramatic increase of working depth such that spatial resolution of 1.5 μm can be maintained up to 11 MFPs. The achievable imaging depth is reduced with the increase of target spatial resolution due to the low-pass filtering operation of imaging system. For example, the spatial resolution of 0.6 μm can be maintained only up to 8 MFPs at the same condition. However, the specimen-induced aberrations in biological tissues hinder the collective accumulation of single-scattered waves such that the achievable depth is even shallower than this fundamental limit by a few MFPs.

On the other hand, methods for dealing with aberration have been actively proposed in the field of adaptive optics[13]. The aberrations used to be characterized on the basis of Zernike polynomials[14] by direct wavefront sensing[15,16] or experimental feedback control[17-20]. These approaches have been particularly useful for fluorescence imaging because only the aberration correction of incident waves matters. Nevertheless, the ability to address both multiple scattering and aberrations has been limited by an insufficient number of control elements in wavefront shaping devices. The adaptive optics for coherent imaging has proved even more difficult to implement when multiple scattering noise exists, and successful implementations have only been reported for cases with negligible multiple light scattering[21].

The limitation of being able to deal with either the scattering or the aberration, not both, is apparent. If the background noise caused by multiple light scattering is not addressed, the intensity of the single-scattered waves is less than the intensity of the background noise caused by the multiple-scattered ones. Therefore, an object image cannot be resolved. On the other hand, if aberration is not addressed, then the single-scattered waves are accumulated so ineffectively that they may not effectively compete with the multiple-scattered ones. In this article, we present an experimental method that can address both multiple light scattering and sample-induced aberrations at the same time. The CASS microscopy previously developed was upgraded in such a way as to take the time-resolved amplitude and phase maps of the reflected waves over random illumination patterns across an increased numerical aperture. In the image reconstruction process, we introduced separate angle-dependent phase corrections for the incident and reflected waves, and identified phase corrections in such a way to preferentially accumulate single-scattered waves over multiple-scattered ones for both the forward and phase-conjugation processes. Using this what we term 'closed-loop accumulation of single scattering' (CLASS) microscopy, we could independently correct the aberrations of single-scattered waves on the way to and from a target object even in the presence of strong multiple-scattered waves. With this novel CLASS microscopy, we could not only optimize the accumulation of single-scattering, but also significantly reduce the effect of image distortion. In doing so, we achieved a spatial resolution of 600 nm up to the imaging depth of 7 scattering mean free paths, an unprecedented imaging depth at such a high resolving power.

**The effects of sample-induced aberrations in the presence of strong multiple scattering**

Let us consider a plane wave, $E(x, y, z = 0; \vec{k}^i) = \exp[-ik_x^i x - ik_y^i y]$, incident to a target object embedded in a thick scattering medium, where $\vec{k}^i = (k_x^i, k_y^i)$ is the transverse wavevector of the incident wave (Fig. 1a). When this wave travels through the scattering medium of thickness $L$, the intensity of the wave that preserves its original momentum is attenuated by a factor of $\exp(-L/l_s)$, where $l_s$ is the scattering mean free path, due to multiple light scattering. Moreover, this unscattered wave undergoes the phase retardation $\phi_i(\vec{k}^i)$ depending on $\vec{k}^i$. This unscattered wave is then reflected by the target object whose amplitude reflectance can be described by the object function $O(x,y)$, and gains the transverse wavevector $\Delta \vec{k}$ driven by the object spectrum $\mathcal{O}(\Delta \vec{k})$, which is the Fourier transform of the object function. On its way out, the wave that now has the wavevector of $\vec{k}^o = \vec{k}^i + \Delta \vec{k}$ is again attenuated by the multiple scattering process and also experiences the additional aberration described by the angle-dependent phase retardation $\phi_o(\vec{k}^o)$. Therefore, the angular spectrum of the reflected wave that has the flight time of $\tau_0 = 2L/c$ is written as

$$\mathcal{E}_o(\vec{k}^o; \vec{k}^i) = \sqrt{\gamma} P_o^a(\vec{k}^i + \Delta \vec{k}) \mathcal{O}(\Delta \vec{k}) P_i^a(\vec{k}^i) + \sqrt{\beta} \mathcal{E}_o^M(\vec{k}^i + \Delta \vec{k}; \tau_0). \qquad (1)$$

Here the first term on the right-hand side is the single-scattered wave, and the second term is the multiple-scattered waves that have the same wavevector and flight time as those of the single-scattered wave. The remaining multiple-scattered waves can be ruled out by time-gated detection[12]. $P_i^a(\vec{k}^i) = P(\vec{k}^i) \exp[-i\phi_i(\vec{k}^i)]$ and $P_o^a(\vec{k}) = P(\vec{k}^o) \exp[-i\phi_o(\vec{k}^o)]$ are the complex pupil functions for the illumination and reflection paths, respectively, where $P(\vec{k})$ is the pupil function of the ideal objective lens ($P(\vec{k}) = 1$ for $|\vec{k}| \leq k_0 \alpha$ with $\alpha$ the numerical aperture of the objective lens and $k_0$ the magnitude of the wavevector in free space, and otherwise $P = 0$). The factor $\gamma = \exp[-2L/l_s]$ describes the intensity attenuation of the single-scattered wave for the round trip through the target. $\beta$ is the average intensity of the multiple-scattered waves detected at the camera, which is determined by the imaging optics, the time-gating window, and the optical properties of the scattering medium. The single-scattered wave, which contains the object information, can be obscured by strong multiple-scattered waves when $\gamma/\beta$ is reduced with increasing target depth.

The CASS microscopy that we previously reported was designed to deal with strong multiple scattering backgrounds by the collective accumulation of single-scattered waves. Suppose that we measure the spatial frequency spectra of the reflected waves for $N_m$ different incident wavevectors. In order to preferentially accumulate the single-scattered waves, those reflected waves that originate from the same object spectrum $\Delta \vec{k}$ are coherently added. This is mathematically expressed as

$$\mathcal{E}_{CASS}(\Delta \vec{k}) = \sum_{\vec{k}^i} \mathcal{E}_o(\vec{k}^i + \Delta \vec{k})$$

$$= \sqrt{\gamma} \mathcal{O}(\Delta \vec{k}) \cdot \sum_{\vec{k}^i} P_i^a(\vec{k}^i) P_o^a(\vec{k}^i + \Delta \vec{k}) + \sqrt{\beta} \sum_{\vec{k}^i} \mathcal{E}_o^M(\vec{k}^i + \Delta \vec{k}). \quad (2)$$

This addition process is essential in that the summation at the first term on the right-hand side of Eq. (2), which is a cross-correlation between the complex pupil functions of the illumination and the reflection, amplifies the object function in proportion to $N_m$. In contrast, the amplitude of the multiple-scattered waves grows in proportion to $\sqrt{N_m}$. Therefore, the signal to noise ratio of the intensity is increased from $\gamma/\beta$ to $(\gamma/\beta)N_m$, and the single scattering intensity can outgrow that of multiple scattering when $N_m > \beta/\gamma$.

However, the existence of aberrations can significantly undermine the accumulation of the single scattering signal. The cross-correlation of the complex-valued pupil functions is smaller than that in the aberration-free case due to the following inequality:

$$\left| \sum_{\vec{k}^i} P_i^a(\vec{k}^i) P_o^a(\vec{k}^i + \Delta \vec{k}) \right| \leq \left| \sum_{\vec{k}^i} P(\vec{k}^i) P(\vec{k}^i + \Delta \vec{k}) \right|. \quad (3)$$

Consequently, the single scattering signal that would have been resolved can still be smaller than the multiple scattering signals. Moreover, this cross-correlation adds $\Delta \vec{k}$-dependent phase retardation to the measured object function, thereby distorting the reconstructed object image.

To understand the effect of aberration in the presence of multiple light scattering, we performed numerical simulation for the condition that $\beta = 200\gamma$ and $N_m = 1,245$, which corresponds to the number of free modes for a 20×20 μm² field of view. Also, we introduced the arbitrary aberrations $\phi_i(\vec{k}^i)$ and $\phi_o(\vec{k}^o)$ as depicted in Figs. 1b and 1c, respectively. The amplitude of the cross-correlation map of these two complex pupil functions (Fig. 1d) was well below unity, suggesting that the accumulation of single scattering would be compromised. Figures 1e and 1f show CASS images without and with aberrations, respectively, in the absence of multiple scattering. The target objects were a pair of point particles separated by a distance of 600 nm, which corresponds to the diffraction-limit resolution for 0.8 NA at the source wavelength $\lambda = 800$ nm. As expected, the aberrations made the two particles completely indistinguishable. Figures 1g and 1h show CASS images without and with aberrations, respectively, but this time in the presence of multiple scattering. As long as there is no aberration, CASS microscopy works well even if there is strong multiple scattering (Fig. 1g). The simultaneous presence of scattering and aberration (Fig. 1h) makes it even more difficult to resolve the two particles than in the aberration-only case because, in addition to being improperly accumulated, the single-scattered waves are concealed by the multiple-scattered waves.

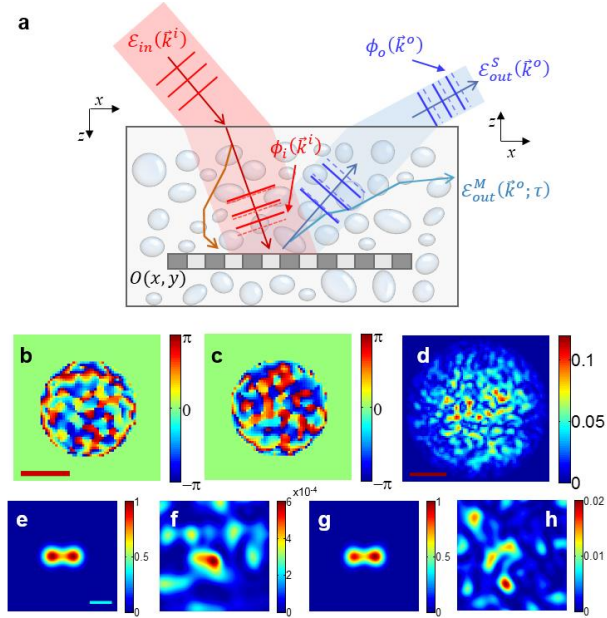

**Figure 1. The effect of sample-induced aberration in imaging targets inside a thick scattering medium. a**, Description of specimen-induced aberrations in the illumination and reflection processes. The phase of the unscattered component of an incident wave with wavevector $\vec{k}^i$ is retarded by $\phi_i(\vec{k}^i)$, and the phase of the reflected wave from the target object is retarded by $\phi_o(\vec{k}^o)$. **b** and **c**, Angle-dependent phase retardations simulated for incident and reflected waves, respectively. For the illumination and reflection processes, random phase shifts were added to the average spherical aberration induced by a 1 mm-thick tissue (refractive index n~1.37) in water. Circular central region corresponds to the pupil of the objective lens with numerical aperture of 0.8. **d**, Amplitude transfer function of CASS microscopy obtained by cross-correlation between the input and output aberrations shown in **b** and **c**. Color scale indicates amplitude transmittance. Scale bars in **b** and **d** correspond to $k_0\alpha$. **e-h**, CASS images of two point particles separated by 600 nm, the diffraction limit of the model system. **e**: without either aberration or multiple scattering. **f**: in the presence of the aberrations shown in **b** and **c**, but with no multiple scattering. **g**: in the presence of multiple scattering but with no aberration. **h**: in the presence of both aberration and scattering. The mean intensity of the multiple scattering was set 200 times stronger than that of single scattering. Color scales in **e-h** are normalized by the peak value in **e**.

**Theoretical framework for dealing with aberrations in the presence of strong multiple light scattering**

We developed a method to find the angle-dependent phase corrections $\theta_i(\vec{k}^i)$ and $\theta_o(\vec{k}^o)$ for the illumination and reflection paths, respectively, to cancel out the respective angle-dependent aberrations, $\phi_i(\vec{k}^i)$ and $\phi_o(\vec{k}^o)$. The main concept of the proposed method is to identify the $\theta_i(\vec{k}^i)$ that would maximize the total intensity of the reconstructed image by preferentially counteracting $\phi_i(\vec{k}^i)$. Then, as an important additional step, we employed the phase conjugation operation and applied $\theta_o(\vec{k}^o)$ to maximize the reconstructed image in the reverse process, which then corrects $\phi_o(\vec{k}^o)$. Repeating these operations leads to the independent identification of $\phi_i(\vec{k}^i)$ and $\phi_o(\vec{k}^o)$.

We first applied initial arbitrary angle-dependent phase corrections $\theta_i^{(1)}(\vec{k}^i)$ to the spectrum of the CASS image to form a CLASS image, i.e. $\mathcal{E}_{CLASS}^{(1)}(\Delta\vec{k}) = \sum_{\vec{k}^i} \mathcal{E}_o(\vec{k}^i + \Delta\vec{k}) e^{i\theta_i^{(1)}(\vec{k}^i)}$. Then we identified the set of $\theta_i^{(1)}(\vec{k}^i)$ that maximizes the total intensity of the CLASS image:

$$\max_{\theta_i^{(1)}(\vec{k}^i)} \sum_{\Delta\vec{k}} \left|\mathcal{E}_{CLASS}^{(1)}(\Delta\vec{k})\right|^2. \quad (4)$$

This can be simply be achieved by changing the individual $\theta_i^{(1)}(\vec{k}^i)$ from 0 to $2\pi$ and finding the particular value of $\theta_i^{(1)}(\vec{k}^i)$ at which the total intensity of the CLASS spectra is maximized. It is important to note that mainly the single-scattered waves take part in this process and the multiple-scattered waves play little role. The maps of multiple-scattered waves taken at different angles of illumination are uncorrelated with respect to one another, and remained so even after multiplying the phase corrections. Therefore, the maximization of the total intensity of the CLASS image is almost exclusively due to the aberration correction of the single-scattered waves.

Through the maximization process, $\theta_i^{(1)}(\vec{k}^i)$ preferentially cancels out the input aberration $\phi_i(\vec{k}^i)$. This becomes evident when examining the cross-term between two representative incident wavevectors, $\vec{k}_1^i$ and $\vec{k}_2^i$, in Eq. (4):

$$\exp i\{\phi_i(\vec{k}_2^i) - \phi_i(\vec{k}_1^i)\} \cdot \exp i\{-\theta_i^{(1)}(\vec{k}_2^i) + \theta_i^{(1)}(\vec{k}_1^i)\}$$

$$\cdot \left[\sum_{\Delta k} \{O(\Delta k) P_o^a(\Delta\vec{k} + \vec{k}_1^i)\}\{O(\Delta k) P_o^a(\Delta\vec{k} + \vec{k}_2^i)\}^*\right]. \quad (5)$$

The term is the cross-correlation between the spectra for the two incident wavevectors indicated in Figs. 2a and 2b. If there was no aberration in the reflection beam path, then $P_o^a = P_o$. The phase angle of the term in the square brackets '[ ]' in Eq. (5), which we define as $\Phi_i^{(1)}(\vec{k}_1^i, \vec{k}_2^i)$, would be zero such that $\theta_i^{(1)}(\vec{k}_2^i) - \theta_i^{(1)}(\vec{k}_1^i) = \phi_i(\vec{k}_2^i) - \phi_i(\vec{k}_1^i)$. Since only relative phase matters, we can set $\theta_i^{(1)}(\vec{k}_1^i) = 0$ and $\phi_i(\vec{k}_1^i) = 0$ at $\vec{k}_1^i = 0$. Then $\theta_i^{(1)}(\vec{k}^i)$ is equal to $\phi_i(\vec{k}^i)$ for arbitrary $\vec{k}^i$, suggesting that the aberration relating from the illumination path is perfectly corrected. In reality, however, aberration also develops through the reflection process, and $\Phi_i^{(1)}(\vec{k}_1^i, \vec{k}_2^i)$ would be nonzero and consequently act as the error for the aberration correction.

For the aberration correction to be effective, the correction error $\Phi_i^{(1)}(\vec{k}_1^i, \vec{k}_2^i)$ should have a finite width of distribution around zero, not a random and uniform distribution between $-\pi$ and $+\pi$. This becomes possible when $P_o^a(\Delta \vec{k} + \vec{k}^i)$ is a slowly varying function with respect to $\Delta \vec{k}$. If we record individual images over a wide view field, then the spectral resolution of the individual complex field images, which is the reciprocal of the width of view field, can be fine enough to make $P_o^a$ a slowly varying function. Figure 2c shows the distribution of $\Phi_i^{(1)}(\vec{k}_1^i, \vec{k}_2^i)$ for the severe aberrations considered in Figs. 1b and 1c for a view field of 20×20 μm², for which it was observed that the width of distribution was indeed finite. In the end the maximization process in Eq. (5) will lead to $\theta_i^{(1)}(\vec{k}^i) \approx \phi_i(\vec{k}^i)$ up to the accuracy given by the width of the distribution of $\Phi_i^{(1)}(\vec{k}_1^i, \vec{k}_2^i)$.

We confirmed the effectiveness of this operation using a numerical simulation for the same aberrations considered in Fig. 1. The first image in Fig. 2g shows that $\vec{\theta}_i = \vec{\theta}_i^{(1)}$ after this first round of the maximization process. The $\vec{\theta}_i^{(1)}$ resembles the input aberration shown in Fig. 1b with a correlation value of 35 %. The first image in Fig. 2h shows the CLASS image reconstructed after applying this phase correction. The existence of particles becomes better visualized than before. However, the resolving power has not yet sufficiently recovered to distinguish the two particles because the output aberration has not yet been addressed.

The maximization operation in Eq. (5) is incomplete because only the aberration arising from the incident wave can be dealt with. In order to form a closed-loop correction, we developed a phase conjugation process in which the wave is incident from $-\vec{k}^o$ to $-\vec{k}^i = -(\vec{k}^o - \Delta\vec{k})$ (Fig. 2d). This reverse process does not require further data acquisition because it can be computed from the set of images originally measured by the reciprocity of wave propagation. Through this reverse process, the CLASS spectrum can be reconstructed as

$$\mathcal{E}^{pc}_{CLASS}(\Delta\vec{k}) = \sqrt{\alpha}\mathcal{O}^*(\Delta\vec{k}) \cdot \sum_{\vec{k}^o} P_o^a(\vec{k}^o)^* P_i^a(\vec{k}^o - \Delta\vec{k})^* + \sqrt{\beta}\sum_{\vec{k}^o} \mathcal{E}_o^M(\vec{k}^o - \Delta\vec{k})^*. \quad (6)$$

Note that the summation operation is now performed over $\vec{k}^o$. With the correction $\vec{\theta}_i = \vec{\theta}_i^{(1)}$ in place, as indicated by the use of a virtual spatial light modulator in Fig. 2d, we apply the phase correction to the output and identified the $\vec{\theta}_o = \vec{\theta}_o^{(1)}$ that would maximize the total intensity of the phase-conjugated CLASS image. Similar to the correction for the illumination path, this iteration leads to the convergence of $\vec{\theta}_o^{(1)}(\vec{k}^o)$ to $\phi_o(\vec{k}^o)$ with the error of correction $\Phi_o^{(1)}(\vec{k}_1^o, \vec{k}_2^o)$ given by the correlation of the two representative spectra shown in Fig. 2e. In fact, this correction for the reflection process converges faster than that of $\vec{\theta}_i^{(1)}$ to $\phi_i(\vec{k}^i)$. Because of the prior correction $\vec{\theta}_i^{(1)}$, the width of the phase histogram of $\Phi_o^{(1)}(\vec{k}_1^o, \vec{k}_2^o)$ (Fig. 2f) is narrower than that of $\Phi_i^{(1)}(\vec{k}_1^i, \vec{k}_2^i)$ (Fig. 2c).

The second image in Fig. 2g shows the $\vec{\theta}_o^{(1)}$ identified through this phase-conjugation process, and it shows good correlation with the original aberration map in Fig. 1c with a correlation value of about 66 %. After this first round of illumination and reflection corrections, the reconstructed CLASS image shown in the second image of Fig. 2h now resolves the two particles better than before. Since the identified aberration maps are not yet complete, we iterated the aberration correction to improve its accuracy. We observed that the accumulated phase corrections converge to the system aberrations as the number of iterations $n$ is increased, i.e. $\sum_n \theta_i^{(n)}(\vec{k}^i) \to \phi_i(\vec{k}^i)$ and $\sum_n \theta_o^{(n)} \to \phi_o(\vec{k}^o)$ (Fig. 2g). For this particular example, 3 rounds of iteration led to the determination of the input and output aberrations to an accuracy of 97 %. The reconstructed image becomes almost the same as the ideal image shown in Fig. 1e. The resulting CLASS images show the increase in the signal intensity, suggesting that the cross-correlation of the aberration-corrected pupil functions had been increased in magnitude. Taken together, these observations confirm that the proposed method works extremely well, even in the presence of strong multiple-scattered waves.

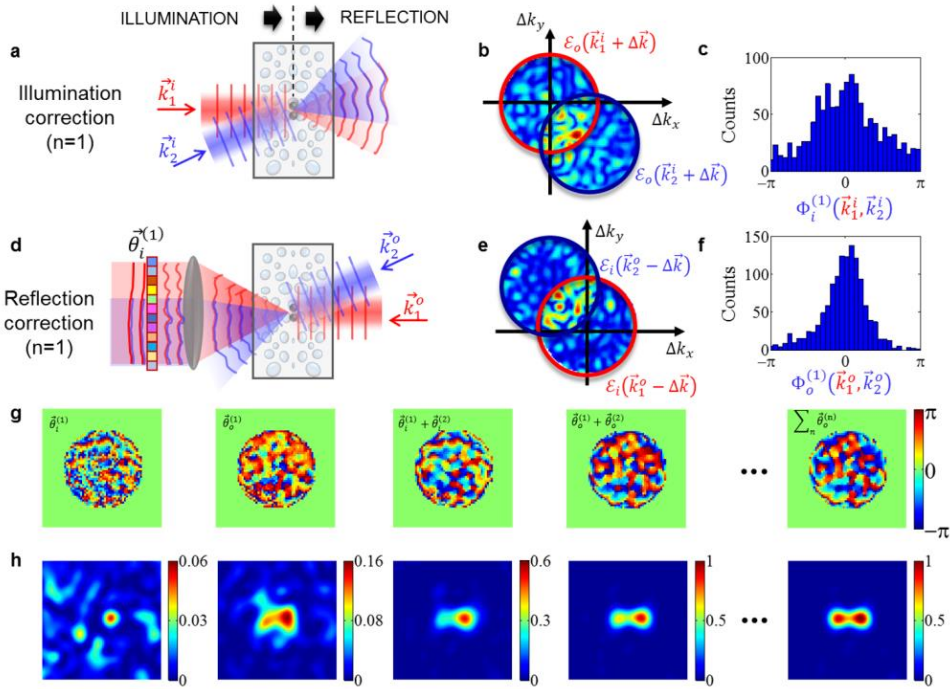

**Figure 2. Algorithms for the independent identification of aberrations for the illumination and reflection paths in the presence of strong multiple light scattering.** Aberrations and the intensity ratio of single- to multiple-scattered waves are the same as those considered in Fig. 1. **a,** Comparison of the wave propagations for the illumination of the two representative incident wavevectors, $\vec{k}_1^i$ and $\vec{k}_2^i$. For the sake of simplicity, the reflection process is drawn to the right-hand side of the illumination process. **b,** Superposition of the two reflected waves shown in **a** in the $\Delta\vec{k}$ space. The radius of each circle is $k_0\alpha$. **c,** The histogram of the phase correction error $\Phi_i^{(1)}(\vec{k}_1^i, \vec{k}_2^i)$ calculated from the superposition of the two reflected waves shown in **b**. The phase angles were calculated for all $\vec{k}_2^i$ while $\vec{k}_1^i$ is fixed at $\vec{k}_1^i = 0$. **d,**

Similar to **a**, but for the phase-conjugated process. Two arbitrarily chosen waves with wavevectors, $\vec{k}_1^o$ and $\vec{k}_2^o$, are considered to be incident from the reflection side. The correction $\vec{\theta}_i = \vec{\theta}_i^{(1)}$ is applied in the data processing, and indicated as the virtual spatial light modulator (vSLM) at the focal plane of the objective lens. **e,** Superposition of the two reflected waves shown in **d** in the $\Delta\vec{k}$ space. **f,** The histogram of the phase correction error $\Phi_o^{(1)}(\vec{k}_1^o, \vec{k}_2^o)$ calculated from the superposition of the two reflected waves shown in **e**. The phase angles were calculated for all $\vec{k}_2^o$ while $\vec{k}_1^o$ is fixed at $\vec{k}_1^o = 0$. **g,** Angle-dependent phase corrections $\vec{\theta}_i^{(n)}$ and $\vec{\theta}_o^{(n)}$ as the iteration number $n$ is increased. Scale bar corresponds to $k_0\alpha$. Color bar, phase in radians. **h,** Reconstructed CLASS images when $\vec{\theta}_i^{(n)}$ and $\vec{\theta}_o^{(n)}$ in **g** were applied. Color bar, intensity normalized by the peak value of the reconstructed image obtained at the end of iteration. Total number of iterations was 6 for the presented data.

**Experimental demonstration of CLASS microscopy with phantom samples**

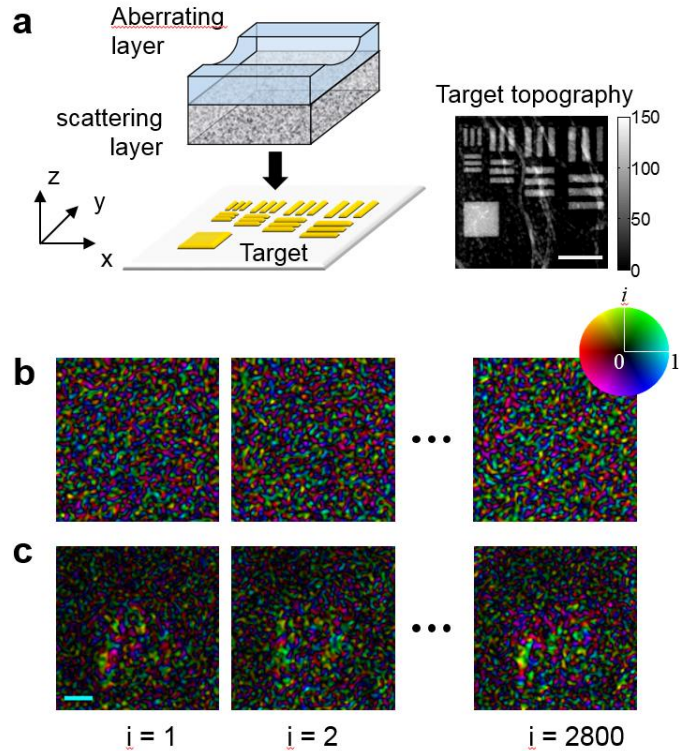

**Figure 3. Experimental measurements of a time-resolved reflection matrix. a**, Layout of the phantom sample geometry. An asymmetric aberrating layer made of a clean PDMS block with a dent in the form of a cylindrical groove was placed on the top of a $7l_s$-thick scattering layer. A resolution target was placed underneath the scattering layer. The topography of the target measured by atomic force microscopy is shown

in gray scale. Scale bar, 4 μm. Color map, height in nanometers. **b**, Complex field maps of the incident waves generated by writing random phase patterns on the SLM. Images were acquired by placing a clean mirror at the sample stage. Only a few representative images from 2,800 are shown. **c**, Complex field maps taken after placing the target shown in **a** at the sample stage and corresponding to the same set of illumination patterns used in **b**. Scale bar, 5μm. Images in **b** and **c** are normalized by their respective maximum amplitudes. The color map indicates both the amplitude and phase of the complex field.

To experimentally demonstrate CLASS microscopy, we upgraded the previous version of the CASS microscope[12] in two major respects. First, a water-dipping type objective lens (Nikon, CFI-Apo-40XW-NIR, 0.8 NA) with a twice larger numerical aperture than before was used for sub-micron resolution imaging. The water-dipping feature helps to minimize back reflection from the surfaces of biological tissues and specimen-induced spherical aberrations. Second, as the basis of illumination we used a set of random patterns composed of multiple plane waves rather than individual angular plane waves. Interferometric imaging has uncontrolled phase shifts due to the mechanical fluctuation of the relative path length between the sample and reference waves. When using plane-waves, these phase shifts are indistinguishable from the angle-dependent phase retardation caused by the sample, meaning that the sample-induced aberrations cannot be identified. When using random patterns, however, multiple plane waves of known relative phases are simultaneously injected to the sample such that we can exclusively deal with sample-induced phase shifts.

For a controllable test target, we prepared an asymmetric aberrating layer featuring a cylindrical groove along the *y* direction with a radius of curvature of 6.0 mm (Fig. 3a). The layer was made of 1.0 mm-thick clean PDMS (refractive index *n*=1.41). Because of the refractive index mismatch between the layer and the immersion medium (water, *n*=1.33), the cylindrical groove causes asymmetric aberrations such as astigmatisms. A $7l_s$-thick scattering layer was placed underneath this aberrating layer. This arrangement allowed aberration and scattering to be controlled independently. These layers were placed on the top of a resolution target made by focused-ion-beam (FIB) milling of a gold-coated slide glass (gray scale image in Fig. 3a). The finest lines have a separation of 600 nm.

We measured the time-resolved reflection matrix of this test target and later that of biological tissues. Random phase patterns numbering 2,800 were sequentially written on a spatial light modulator (SLM, Hamamatsu X10468) located at the conjugate plane of the sample in order to cover all the orthogonal free modes determined by the illumination area of $30 \times 30 \ \mu m^2$ and the spatial frequency bandwidth corresponding to 0.8 NA. The total acquisition time for the entire set of images was about 5 minutes but could potentially be reduced to a few seconds if high-speed SLM and camera were used. Then we recorded a time-resolved complex field map $u_i(x, y; j, \tau)$ for the $j^{th}$ random pattern written on the SLM by placing an ideal mirror at the sample plane for the arrival time $\tau$ associated with the depth of the mirror. The representative complex-field maps shown in Fig. 3b exhibit speckle patterns due to the use of random patterns of illumination. For the same set of illumination patterns, the complex-field map $u_o(x, y; j, \tau)$ of the reflected waves (Fig. 3c) was recorded by placing the scattering medium shown in Fig. 3a at the sample stage.

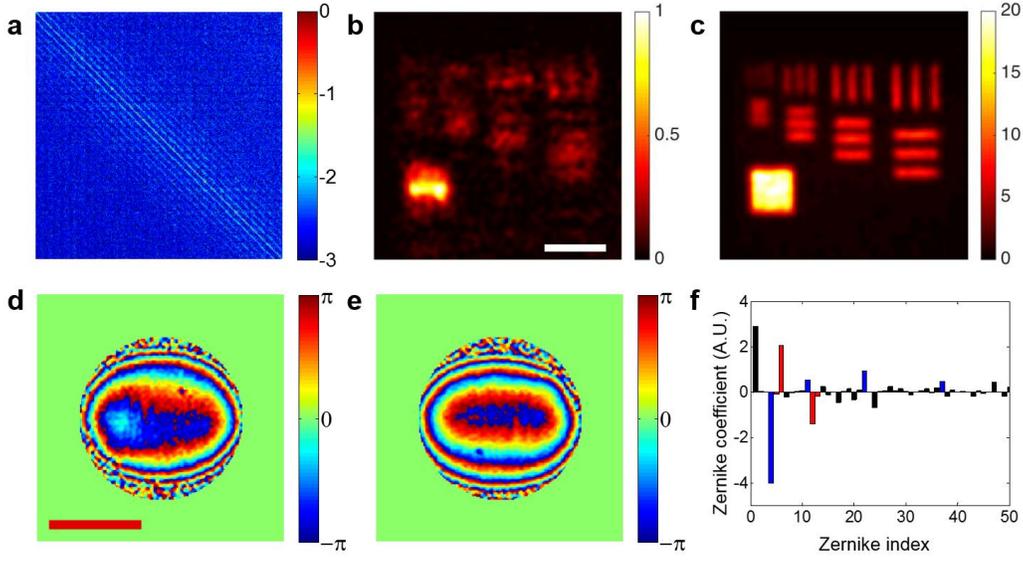

**Figure 4. Experimental demonstration of the addressing of both scattering and aberration using CLASS microscopy. a**, Reconstructed time-resolved reflection matrix, $\mathcal{E}(\vec{k}^o; \vec{k}^i)$ in the spatial frequency domain. Column index indicates reflected wavevector and row index indicates incident wavevector. Only a part of the matrix ($|k| < 0.26k_0$) is shown in the figure. Color scale, log scaled amplitude. **b-c**, CLASS images before and after the application of aberration corrections, respectively. Color bars, intensity normalized by the maximum intensity in **b**. Scale bar, 5μm. **d-e**, Angle-dependent phase corrections, $\theta_i(\vec{k}^i)$ and $\theta_o(\vec{k}^o)$, respectively, identified from the experimental data. Color bars, phase in radians. Scale bar, $k_0\alpha$. **f**, Decomposition of the map, $\theta_o(\vec{k}^o)$, in e into the first 50 Zernike polynomials $Z_j$ following the convention of Noll's sequential indices. Blue bars indicate spherical aberrations ($j = 4, 11, 22, 37$), and red bars astigmatisms ($j = 5, 6, 12, 13$).

By taking the Fourier transform of the set of images in Figs. 3b and 3c, we constructed the matrices describing the incident and reflected waves, $\mathcal{U}_i(\vec{k}^i; i)$, and $\mathcal{U}_o(\vec{k}^o; i)$, respectively. From the product $\mathcal{U}(\vec{k}^o; i) \cdot \mathcal{U}_i^{-1}(\vec{k}^i; i)$, we could construct the time-resolved reflection matrix $\mathcal{E}_o(\vec{k}^o; \vec{k}^i)$, which is shown in Fig. 4a. In the figure, we rearranged the two-dimensional spatial frequencies $(k_x, k_y)$ into single columns of matrix. Therefore, each column corresponds to the output spectrum $\mathcal{E}_o(\vec{k}^o; \vec{k}^i)$ shown in Eq. (1) taken for a given incidence wavevector, $\vec{k}^i$. This procedure allowed the conversion of the initial random speckle basis into the basis of spatial frequency.

To the time-resolved reflection matrix shown in Fig. 4a, we applied Eq. (2) to construct an ordinary CASS image (Fig. 4b). For this particular sample, the time-gated multiple scattering intensity was about 10 times larger than that of the single scattering intensity ($\beta = 10\gamma$). Therefore, the structures would have been resolved with $N_m$ = 2,800 measurements if there were no aberrations. Due to the pronounced sample-induced aberrations, however, the target structures, especially fine structures, were invisible. However, by using CLASS microscopy, we could dramatically improve the spatial resolution and signal strength of the single-scattered waves (Fig. 4c). (see Methods for the detailed correction procedure). The finest structures with a diffraction-limited line spacing of 600 nm were clearly resolved. Moreover, through the aberration correction the magnitude of the single scattering intensity was increased by about 200 times (See SI for the detailed analysis). This confirms that the proposed method could successfully address both scattering and aberration up to the full system numerical aperture.

The acquired angle-dependent phase corrections for the illumination and reflection paths are shown in Figs. 4d and 4e, respectively. Because of the asymmetric aberration along the y-direction caused by the cylindrical groove, phase corrections along the $k_y$ direction have much steeper variations than those along the $k_x$ direction. In order to quantify the obtained aberration, we decomposed the output phase correction map in Fig. 4e into Zernike polynomials ($Z_j$), which are widely used to explain these types of aberrations. Because the aberration correction of CLASS microscopy was performed at the spatial frequency resolution of 1/30 $\mu m^{-1}$, we could identify the high order Zernike modes that ordinary adaptive optical microscopy cannot. In Fig. 4f we show only the first 50 coefficients according to Noll's sequential indices[22]. As expected, the dominant components were the astigmatism ($j$=5, 6, 12, and 13, red bars) induced by the cylindrical aberrating layer and the spherical aberration ($j$=4, 11, 22, 37, blue bars) induced by the overall index mismatch between the immersion medium and the scattering layers. This result confirms that the proposed method can successfully identify sample-induced aberrations even in the presence of strong multiple light scattering.

**Ultra-high resolution imaging of targets underneath biological tissues**

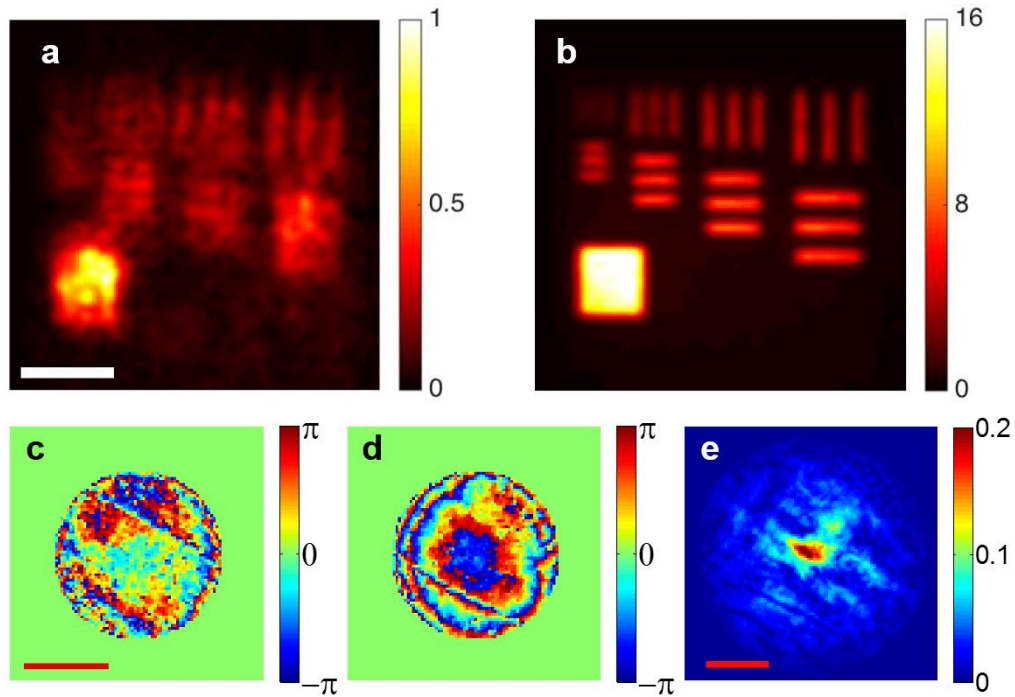

**Figure 5. Experimental demonstration of aberration correction for a target underneath fresh rat brain tissue. a** and **b**, CLASS images of a resolution target under a 500μm-thick rat brain tissue layer before and after aberration correction, respectively. Scale bar, 5 μm. Color bars, intensity normalized by the maximum intensity in **a**. **c** and **d**, Input and output phase correction maps, respectively. Scale bar, $k_0\alpha$. Color bars, phase in radians. **e**, Amplitude transfer function calculated by cross-correlation between **c** and **d**. Scale bar, $k_0\alpha$.

In general, biological tissues exhibit much more complicated aberrations than the phantoms, making high-resolution imaging even more difficult. We demonstrated the performance of CLASS microscopy on targets located under a layer of fresh biological tissue. A 500 μm-thick slice of rat brain tissue, whose scattering mean free path was measured to be approximately 100 μm, was placed on the top of the resolution target. As shown in Fig. 5a, the CLASS image without aberration correction could not reveal the fine structures of the resolution target due to the multiple scattering noise and the aberration induced by the tissue. On the other hand, when full aberration correcting functionality of CLASS microscopy was used, the targets were clearly visible up to the line spacing of 600 nm. Also, the intensity fluctuation in the gold coated area was remarkably reduced, supporting the conclusion that the aberration correction process properly accumulated the single-scattered waves.

The angle-dependent phase corrections for illumination and reflection identified for the biological tissues are shown in Figs. 5c and 5d, respectively. Unlike the results shown in Figs. 4d and 4e, irregular patterns appeared due to the complex internal structures in the brain tissue. In Fig. 5e, we present the amplitude transfer function of the rat brain tissue created by calculating the cross-correlation of input and output aberration maps. The color scale is normalized by the maximum value of the ideal amplitude transfer function of CASS microscopy. The attenuated value of the amplitude transfer function and reduced bandwidth were responsible for the deterioration of the ordinary CASS image in Fig. 5a. In addition, while the phase gradient of aberration is relatively flat at low spatial frequency, it steepens at higher spatial frequencies. This difference underlies the necessity of compensating for specimen-induced aberrations in ultra-high resolution imaging.

**Discussion**

We presented an experimental method that can perform sub-micron resolution imaging of targets located up to the depth of 7 scattering mean free paths. The proposed method introduced separate angle-dependent phase corrections for the illumination and reflection paths to preferentially optimize the total intensity of single-scattered waves in both forward and phase-conjugation processes. Our method is unique in four major respects. First, the optimization operation of the total intensity in the momentum difference space acted mainly on the single-scattered waves, rather than the multiple-scattered ones. For this reason, the proposed method could be successful even with strong multiple scattering backgrounds. Second, due to the use of the closed-loop operations, aberrations for the illumination and reflection paths could be independently identified without the need for an embedded ideal point source, often called a guide[23]. Unlike fluorescence imaging where the correction of aberrations from the illumination path is the only concern, this capability is especially critical for coherent imaging as aberrations from both paths are responsible for the reduced SNR. The third important aspect is that aberration correction is performed over a wide field of view with finely stepped illumination angles. This enables our method to eliminate the steep angle-dependent phase retardations that thick biological tissues induce at large propagating angles, and thereby outperform conventional adaptive optical microscopy. Finally, the aberration correction is performed in post-processing after the acquisition of the time-resolved reflection matrix. Given the same number of angular bases to correct, this post processing step is much faster than the experimental feedback iteration required for wavefront control. The speed of the current implementation is largely limited by the speed of the SLM, but the use of a high-speed binary control SLM with a high-speed camera would substantially speed up measurement.

The ability to perform ultra-high spatial resolution imaging deep within scattering media will open new possibilities for studying important biological reactions in great detail. Future applications of this technology could include early disease diagnosis, studies of nervous systems and of the activities of stem cells inside bone marrow, and so on. The proposed method can be extended to other coherent imaging modalities such as second-harmonic generation microscopy and stimulated Raman scattering microscopy to increase their imaging depth limits in a similar way. Moreover, this new CLASS microscopy can offer specimen-induced aberration maps with minimal photo-bleaching to various confocal fluorescence imaging microscopy techniques[15]. Ultimately, our study will widen the scope of the applications that optical microscopy can explore.

**Methods**

The data processing procedure for the experimentally measured time-resolved reflection matrix is summarized. We rearranged the elements of the matrix constructed in Fig. 4a to form a matrix $\mathcal{E}_o(\vec{k}^o - \vec{k}^i; \vec{k}^i)$ (Fig. 6b) whose column and row indices are $\vec{k}^i$ and $\vec{k}^o - \vec{k}^i$, respectively. By coherently adding elements of the same rows, a process equivalent to applying Eq. (2), a conventional CASS image can be obtained. We then applied angle-dependent phase correction $\theta_i(\vec{k}^i)$ to each column in Fig. 6b and performed the maximization operation described in Eq. (4). After applying the $\theta_i(\vec{k}^i)$ (Fig. 6c) thus identified to the matrix in Fig. 6b, we rearranged the matrix to form a matrix $\mathcal{E}_o(\vec{k}^o; \vec{k}^i - \vec{k}^o)$ whose column and row indices are now $\vec{k}^i - \vec{k}^o$ and $\vec{k}^o$, respectively. Adding elements in the same columns correspond to reconstructing the CLASS image in the phase-conjugation process (Eq. (6)). Next, we applied $\theta_o(\vec{k}^o)$ to the rows of the matrix to maximize the total intensity of the phase-conjugated CLASS image. After applying $\theta_o(\vec{k}^o)$ thus identified (Fig. 6e) to the matrix in Fig. 6d, the next round of iteration was initiated.

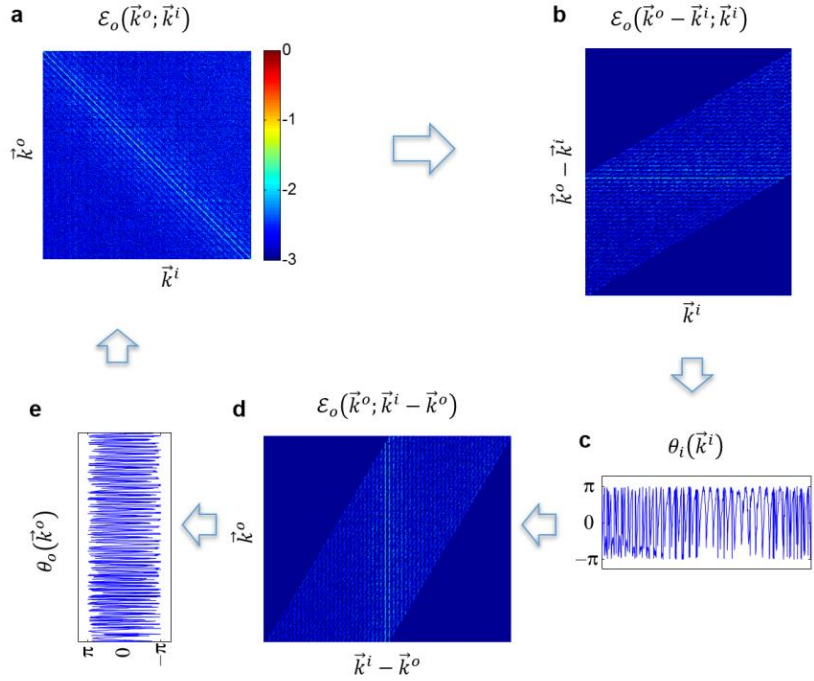

**Figure 6. Flowchart of CLASS algorithm from the experimentally measured TRRM.** For simplicity, a single iteration is presented. **a,** Initial matrix, $\mathcal{E}_o(\vec{k}^o; \vec{k}^i)$ shown in Fig. 4a. **b**, The matrix reshaped to $\mathcal{E}_o(\vec{k}^o - \vec{k}^i; \vec{k}^i)$ **c,** Identifying the $\theta_i(\vec{k}^i)$ that maximizes the coherent summation along the same $\vec{k}^o - \vec{k}^i$, i.e. row direction in **b**. **d**, After applying the $\theta_i(\vec{k}^i)$ identified in **c**, the matrix reshaped to $\mathcal{E}_o(\vec{k}^o; \vec{k}^i - \vec{k}^o)$. **e**, Identifying the $\theta_o(\vec{k}^o)$ that maximizes the coherent summation along the same $\vec{k}^i - \vec{k}^o$, i.e. column direction in **d**.

**References**

1   Ntziachristos, V. Going deeper than microscopy: the optical imaging frontier in biology. *Nat Meth* **7**, 603-614 (2010).
2   Hell, S., Reiner, G., Cremer, C. & Stelzer, E. H. Aberrations in confocal fluorescence microscopy induced by mismatches in refractive index. *Journal of microscopy* **169**, 391-405 (1993).
3   Wyant, J. C. & Creath, K. Basic wavefront aberration theory for optical metrology. *Applied optics and optical engineering* **11**, 2 (1992).
4   Hee, M. R., Swanson, E. A., Izatt, J. A., Jacobson, J. M. & Fujimoto, J. G. Femtosecond transillumination optical coherence tomography. *Opt. Lett.* **18**, 950-952, doi:10.1364/OL.18.000950 (1993).
5   Huang, D. *et al.* Optical Coherence Tomography. *Science* **254**, 1178-1181 (1991).
6   Dubois, A., Vabre, L., Boccara, A.-C. & Beaurepaire, E. High-resolution full-field optical coherence tomography with a Linnik microscope. *Applied optics* **41**, 805-812 (2002).

7	Izatt, J. A., Swanson, E. A., Fujimoto, J. G., Hee, M. R. & Owen, G. M. Optical coherence microscopy in scattering media. *Opt. Lett.* **19**, 590-592 (1994).
8	Kino, G. S. & Corle, T. R. *Confocal scanning optical microscopy and related imaging systems*.   (Academic Press, 1996).
9	Ralston, T. S., Marks, D. L., Carney, P. S. & Boppart, S. A. Interferometric synthetic aperture microscopy. *Nature Physics* **3**, 129-134 (2007).
10	Popoff, S. M. *et al.* Exploiting the Time-Reversal Operator for Adaptive Optics, Selective Focusing, and Scattering Pattern Analysis. *Physical Review Letters* **107**, 263901 (2011).
11	Badon, A. *et al.* Smart optical coherence tomography for ultra-deep imaging through highly scattering media. *arXiv preprint arXiv:1510.08613* (2015).
12	Kang, S. *et al.* Imaging deep within a scattering medium using collective accumulation of single-scattered waves. *Nature Photonics* **9**, 253-258 (2015).
13	Tyson, R. K. *Principles of adaptive optics*.   (CRC press, 2015).
14	Born, M., Wolf, E. & Bhatia, A. B. *Principles of Optics: Electromagnetic Theory of Propagation, Interference and Diffraction of Light*.   (Cambridge University Press, 1999).
15	Rueckel, M., Mack-Bucher, J. A. & Denk, W. Adaptive wavefront correction in two-photon microscopy using coherence-gated wavefront sensing. *Proceedings of the National Academy of Sciences* **103**, 17137-17142 (2006).
16	Tao, X. *et al.* Adaptive optics confocal microscopy using direct wavefront sensing. *Opt. Lett.* **36**, 1062-1064 (2011).
17	Booth, M. J., Neil, M. A., Juškaitis, R. & Wilson, T. Adaptive aberration correction in a confocal microscope. *Proceedings of the National Academy of Sciences* **99**, 5788-5792 (2002).
18	Ji, N., Milkie, D. E. & Betzig, E. Adaptive optics via pupil segmentation for high-resolution imaging in biological tissues. *nAture methods* **7**, 141-147 (2010).
19	Marsh, P., Burns, D. & Girkin, J. Practical implementation of adaptive optics in multiphoton microscopy. *Optics Express* **11**, 1123-1130 (2003).
20	Park, J.-H., Sun, W. & Cui, M. High-resolution in vivo imaging of mouse brain through the intact skull. *Proceedings of the National Academy of Sciences* **112**, 9236-9241, doi:10.1073/pnas.1505939112 (2015).
21	Sincich, L. C., Zhang, Y., Tiruveedhula, P., Horton, J. C. & Roorda, A. Resolving single cone inputs to visual receptive fields. *Nat Neurosci* **12**, 967-969, doi:http://www.nature.com/neuro/journal/v12/n8/suppinfo/nn.2352_S1.html (2009).
22	Noll, R. J. Zernike polynomials and atmospheric turbulence*. *JOsA* **66**, 207-211 (1976).
23	Hardy, J. W. *Adaptive optics for astronomical telescopes*.   (Oxford University Press on Demand, 1998).